# Nanoscale imaging of the electronic and structural transitions in vanadium dioxide

M. M. Qazilbash[1,*], A. Tripathi[1], A. A. Schafgans[1], Bong-Jun Kim[2], Hyun-Tak Kim[2,3], Zhonghou Cai[4], M. V. Holt[5], J. M. Maser[5], F. Keilmann[6], O. G. Shpyrko[1], and D. N. Basov[1]

[1] *Department of Physics, University of California, San Diego, La Jolla, California 92093, USA.*

[2] *Metal-Insulator Transition Center, Electronics and Telecommunications Research Institute (ETRI), Daejeon 305-350, Korea.*

[3] *School of Advanced Device Technology, University of Science and Technology, Daejeon, 305-333, Korea.*

[4] *Advanced Photon Source, Argonne National Laboratory, Argonne, Illinois 60439, USA.*

[5] *Center for Nanoscale Materials, Argonne National Laboratory, Argonne, Illinois 60439, USA.*

[6] *Max Planck Institute of Quantum Optics – Munich Centre for Advanced Photonics and Center for NanoScience, 85748 Garching, Germany.*

**Abstract**

**We investigate the electronic and structural changes at the nanoscale in vanadium dioxide ($VO_2$) in the vicinity of its thermally driven phase transition. Both electronic and structural changes exhibit phase coexistence leading to percolation. In addition, we observe a dichotomy between the local electronic and structural transitions. Nanoscale x-ray diffraction reveals local, non-monotonic switching of the lattice structure, a phenomenon that is not seen in the electronic insulator-to-metal transition mapped by near-field infrared microscopy.**

# I. INTRODUCTION

The interplay between electronic and lattice degrees of freedom is commonplace in correlated electron systems. In recent years, local scanning probes have revealed electronic nanoscale phase-separation in such systems[1,2,3,4,5]. However, entangled structural and electronic effects have never been investigated in tandem at length scales characteristic of the phase-separated materials. A prototype of such systems is the correlated material vanadium dioxide ($VO_2$) whose thermally-driven phase transition at $T_C \sim 340K$ involves dramatic changes in electronic and structural properties. The structural phase transition (SPT) in $VO_2$ between the rutile and monoclinic ($M_1$) lattice structures is associated with dimerization of all the vanadium ions (Peierls instability). Historically, the electronic insulator-to-metal transition (IMT) had been thought to occur due to the Peierls instability caused by coupling of electrons to a soft-phonon mode[6]. However, the discovery of the monoclinic and insulating $M_2$ phase of $VO_2$ highlighted the role of electronic correlations. This is because in the $M_2$ phase, half of the vanadium ions are not dimerized but instead form equally spaced $\mathbf{s} = 1/2$ Heisenberg chains[7].

More recent work on the IMT has revealed fingerprints of significant electron-electron interactions. These fingerprints include: changes in optical spectral weight over energy scales of at least 6 eV [8]; the large energy separation between the different $t_{2g}$-bands in the $M_1$ insulating phase[8,9]; the observation of metallicity in $VO_2$ without the structural transformation to the rutile phase[10,11,12]; the bad metal behavior of the rutile phase[13]; the relationship between $T_C$ and the equilibrium carrier density in the $M_1$ insulator[11,14]; the observation of metallicity in the monoclinic structure in high pressure measurements[15]; the fast timescales for the appearance of metallic conductivity compared to the lattice dynamics[16] ; the observation of the lower Hubbard band in resonant photoemission spectroscopy[17]; and the enhanced optical mass of the charge carriers in the metallic puddles near the IMT . While the characteristics of the IMT in $VO_2$ comply with litmus tests for a Mott transition which occurs as a consequence of strong Coulomb repulsion between electrons, the role of the Peierls instability remains unsettled. Scanning probe techniques have demonstrated that the IMT in $VO_2$ proceeds on nanometer length scales and results in a spatially heterogeneous mixture of coexisting electronic phases in the phase transition regime[2,18,19,20,21]. Therefore, a credible inquiry into the role of the Peierls instability in the electronic IMT in $VO_2$ requires a focus on the evolution of local structural properties near the IMT and SPT.

In this work, we focus on local structural changes in $VO_2$ films imaged with nanoscale scanning x-ray diffraction that provides unprecedented 40 nm spatial resolution. We find that the SPT, like the IMT, proceeds in a percolative manner resulting in coexisting crystalline phases. We also find that the SPT exhibits local non-monotonic switching between the two lattice structures, a phenomenon not observed in the monotonic IMT imaged with scanning near-field infrared microscopy (SNIM). This contrasting behavior between the local electronic and structural order parameters suggests nanoscale dichotomy between the IMT and SPT in the phase transition regime of $VO_2$ films.

## II. EXPERIMENTAL METHODS

We performed nanoscale x-ray diffraction measurements to document the structural transition in highly-oriented $VO_2$ films. $VO_2$ films 80 nm thick were grown on $(\bar{1}012)$ oriented sapphire substrates by the sol-gel method[22]. These $VO_2$ films are highly oriented with (200) orientation of the monoclinic ($M_1$) lattice at room temperature. Across the SPT, the $VO_2$ films undergo a structural change to the rutile lattice with (101) orientation in the tetragonal basis (see Fig.1). Two triangular gold electrodes were patterned with photolithography on to the $VO_2$ film for resistance measurements in parallel with the nano-imaging studies. The triangular electrodes are tapered with a 3.5 μm x 3.5 μm channel of $VO_2$ between them as shown in the inset of Fig. 1. We selected a 2 μm x 2 μm area near the tapered ends of the gold electrodes for x-ray imaging based on the highest intensity of the Bragg diffraction peak from $M_1$ $VO_2$.

Nanoscale scanning probe x-ray diffraction measurements were performed at the Hard X-ray Nanoprobe (HXN) beamline[23,24,25] operated by the Center for Nanoscale Materials in partnership with the Advanced Photon Source at Argonne National Laboratory. The recently developed HXN beamline utilizes Fresnel zone plate optics with 24 nm minimum linewidth integrated with an advanced opto-mechanical nanoscale scanning system to provide a hard x-ray beam spot at a landmark 40nm lateral resolution with a position stability of 2-5 nm relative to the sample[23,24,25]. The scanning probe diffraction microscopy measurements were carried out at 10 keV in reflection geometry at two selected momentum transfers giving the highest possible contrast between monoclinic $M_1$ and rutile phases with integrated intensity collected by a single-photon sensitive detector.

In Fig.1 we show the (200) Bragg peak from the monoclinic $M_1$ lattice and the (101) Bragg peak from the rutile lattice. The rutile peak is shifted from the $M_1$ peak which provides us the basis for examining the structural transformation. A Neaspec scanning near-field infrared microscope operating at a wavelength of 10.7 μm and with a lateral resolution of 15 nm[20,26,27] was used to image the electronic transition in a region located within a few micrometers of the area surveyed by the x-ray probe. It was not possible to perform the SNIM measurements simultaneously on the exact same area as the x-ray nano-diffraction experiments. However, temperature-dependent trends in the local electronic and structural transitions can be deduced from the images obtained from the two measurements. The key advantage of our approach is that the relationship between structural and electronic ordering - observed via domain length scale nucleation, fluctuation, and growth - can now be directly compared as a function of the intrinsic order parameters of the system. Temperature dependence of the resistance of the $VO_2$ device was measured simultaneously with both the x-ray and SNIM experiments and was used to correct for systematic uncertainty in the temperature control between the two experiments.

The temperature in the phase transition regime was increased in small discrete steps. Moreover, appropriate feedback parameters of the temperature controller ensured that the temperature did not overshoot the target temperature. The feedback parameters also ensured that the temperature was stable to within 0.05 K of the set temperature while the sample was raster scanned to obtain the images shown in Fig. 2. The thermal drifts of the sample due to thermal expansion of the heating stage were of the order of a tenth of a micrometer per Kelvin. Therefore, we performed x-ray scans at closely spaced temperatures so that there was substantial common area between images taken at adjacent temperatures allowing us to use cross-correlation analysis between images to correct for the thermal drift. Hence, we were able to study the same 1.2 μm x 1.7 μm common area in the x-ray imaging experiments as a function of temperature. In the case of SNIM measurements, thermal drifts were corrected by identifying features in simultaneously acquired topography images.

### III. RESULTS AND DISCUSSION

Fig. 2 shows two-dimensional 1.2 μm x 1.7 μm spatial maps of the x-ray diffraction intensity with the detector set at the scattering angle $2\theta = 29.58^\circ$. The contrast between the Bragg peaks from the low-temperature monoclinic $M_1$ phase and the high-temperature rutile phase is maximized at this angle of the detector (see Fig. 1). As the sample is heated, the lattice structure transforms from the $M_1$ phase to the rutile phase near the SPT, and the intensity of the Bragg peak associated with the $M_1$ phase decreases. The spatial variation of the intensity of the Bragg peak associated with the $M_1$ phase has been obtained at numerous temperatures while thermally heating the $VO_2$ film through the SPT (see Fig. 2). The spatial resolution of the images is 40 nm in the vertical direction and 160 nm in the horizontal direction. Evidently, as the structure transforms from the $M_1$ phase to the rutile lattice, the intensity in certain patches of the scanned area decreases. For example, one can clearly see a spatial variation of Bragg intensity between temperatures of 336 K and 339 K. A patch of lower Bragg intensity nucleates at $T = 336$ K and rapidly progresses across the scanned area with increasing temperature. This phenomenon is qualitatively similar to the percolative IMT seen in scanning near-field infrared microscopy on the same sample discussed below. We also performed complementary raster scans with the detector angle fixed at $2\theta = 29.92^\circ$ to confirm that the $M_1$ structure indeed transforms to rutile structure (see Fig.5 in the appendix). We note here that the image taken at $T = 360$ K in Fig. 2 indicates that the scanned area completely transitions from the $M_1$ into the rutile structure. This is confirmed by a complementary image obtained at $2\theta = 29.92^o$.

Fig. 3 displays images of spatial variation in the scattered near-field infrared amplitude from the $VO_2$ film undergoing the thermally-driven IMT, collected over the same temperature range on the same sample as the x-ray diffraction images. The lateral spatial resolution of the SNIM images is 15 nm. The depth of the film probed by SNIM is discussed in the appendix. Similar to previous works[2,20,28], the images show nucleation of metallic puddles in the insulating host. The newly formed puddles grow, and additional new puddles nucleate with increasing temperature. These puddles eventually connect forming a percolative network in the scanned region. Notice the coexistence of insulating and metallic phases in the phase transition regime. The SPT shows a similar percolative behavior (Fig. 2), and this suggests that there are distinct but coexisting lattice structures in the phase transition regime. Repeated x-ray and SNIM scans of the same area at a fixed temperature show the same patterns of the respective coexisting phases, and changes in these patterns occur only upon an increase in the temperature. Therefore, local heating due to both x-ray and infrared photons can be ruled out.

The general trend in Fig. 2 is that the monoclinic $M_1$ lattice transforms to the rutile lattice in a percolative manner. However, a closer inspection of the x-ray images reveals that nanoscale regions of rutile structure, once formed, switch back to the monoclinic structure despite the increase in temperature. This phenomenon is not an isolated occurrence because it happens thrice in the same measurement while the sample temperature is slowly increased. It is seen in spatial maps framed with black borders in Fig. 2 where the rutile regions (red color) switch back to monoclinic lattice i.e. regions displayed in green color. Actually, the monoclinic and rutile phases coexist at length scales smaller than our spatial resolution in the regions represented by green color. We had also observed nonmonotonic switching of the lattice structure in previous unpublished x-ray nano-imaging experiments that we performed on a similar $VO_2$ film but we wanted to confirm our previous observation. Surprisingly, no such non-monotonic switching phenomenon has been observed in the IMT mapped with SNIM on the same $VO_2$ films used for x-ray nano-imaging (see Fig. 3). In near-field infrared maps we observe that the metallic regions, once formed, persist and grow with increasing temperature.

In order to gain further insight into the relationship between the Peierls instability and the IMT, we plot the fraction of the monoclinic $M_1$ phase in the imaged common area as a function of temperature in Fig. 4. This can be compared to the fraction of the insulating regions plotted in Fig. 4 which is determined from the SNIM images shown in Fig. 3 using a method described in Ref. 28. The fraction of the $M_1$ phase plotted in Fig. 4 is obtained as follows. The integrated x-ray diffraction intensity at $T = 360$ K (rutile phase) is subtracted from the integrated x-ray diffraction intensity for images obtained at various temperatures in the phase transition regime (shown in Fig. 2). This quantity is divided by the difference in the integrated x-ray diffraction intensities for the images obtained at $T = 330$ K ($M_1$) and $T = 360$K (rutile).

Notice the non-monotonic variation of the fraction of the $M_1$ phase with increasing temperature which indicates that parts of the scanned area switch back to the $M_1$ phase after having transformed to the rutile lattice structure. The fraction of insulating $VO_2$, obtained from the near-field infrared images in Fig. 3, decreases monotonically as the temperature is increased as shown in Fig. 4. This monotonic decrease in the insulating fraction (or monotonic increase of the metallic fraction) while heating $VO_2$ is in accord with all of our previous SNIM studies on similar $VO_2$ films[2,20,28] some of which were performed with even smaller temperature steps of 0.2 K. Moreover, it is consistent with the monotonic decrease of the $dc$ resistance of the $VO_2$ film measured between the two gold electrodes (see Fig. 4). We also note here that the monotonic change in our resistance measurements is in accord with other reports on charge transport in micrometer-size $VO_2$ two-terminal structures[29]. Such structures, which ought to be sensitive to small vacillations of the resistance, do not show any non-monotonic behavior in the electronic transport. Evidently, the phenomena we observe are suggestive of a dichotomy between the electronic and structural transformations at the nanoscale. This is further confirmed by the contrasting trends in the derivatives with respect to temperature of the phase fractions in Fig. 4 (see discussion and Fig. 6 in the appendix).

We performed many angular scans around $2\theta = 29.58^o$ of the x-ray diffraction intensity from local 160 nm x 40 nm regions in the phase transition regime. All such scans can be satisfactorily fitted to a linear combination of Bragg peaks from the $M_1$ and rutile phases (see discussion and Fig. 7 in appendix). No obvious signatures of an intermediate structural phase are present above the noise level of our measurements. The absence of an intermediate structural phase in our $VO_2$ film means that the structural transition is locally abrupt and occurs between the $M_1$ and rutile lattice structures. We have also observed that the rutile structure has the tendency to switch back to the $M_1$ structure while the electronic changes (infrared contrast and resistance) proceed monotonically across the phase transition. The monotonic nature of the temperature-induced electronic changes are consistent with other reports on persistent electronic switching seen in $VO_2$ films[30,31]. Based on the contrasting local behavior of the electronic and structural transitions, we hypothesize that the nanoscale electronic and structural properties become distinct from the conventional macroscopic "$M_1$-insulating, rutile-metallic" paradigm over a narrow temperature range within the phase transition. It is possible that the electronic and structural transitions are decoupled. We stress that the conclusion of a dichotomy between electronic and structural transitions at nanometer length scales is in accord with the inferences of area-averaging techniques[32,33]. Simultaneous structural and electronic nano-imaging of the same particular area of the sample will be required in the future for a deeper understanding of the phase transition in $VO_2$. Our findings for the thermally-induced IMT in $VO_2$ are distinct from the observations in the ultra-fast optically-induced transition[34]. It is possible that the end phase and the path to the optically-induced transition are different from those observed in the thermally-induced transition.

## IV. SUMMARY

In summary, we employed two complementary nano-imaging techniques to investigate the thermally driven phase transition in $VO_2$: scanning near-field infrared microscopy which probes local electronic changes, and scanning x-ray nano-diffraction which registers local structural changes. Our results directly show that both electronic and structural transitions evolve in a percolative manner with increasing temperature. However, we also observe nanoscale, non-monotonic switching of the lattice structure, a phenomenon not apparent in the near-field infrared microscopy images of the electronic insulator-to-metal transition in $VO_2$ films. Our approach combining structural and electronic nano-imaging in the bulk of the sample with lateral spatial resolution matching that of the phase separation paves the way to proper characterization of systems with interacting electronic and lattice degrees of freedom.

## ACKNOWLEDGMENTS

M.M.Q. and D.N.B. acknowledge support from the U.S. Department of Energy under Grant No. DE-FG03-00ER45799. O.G.S. and A.T. acknowledge support by U.S. Department of Energy, Office of Science, Office of Basic Energy Sciences, under Contract DE-SC0001805. B.-J.K and H.-T.K. were supported in part by current jump and creative research projects at ETRI. F.K is supported by Deutsche Forschungsgemeinschaft through Cluster of Excellence Munich-Centre for Advanced Photonics. Use of the Center for Nanoscale Materials was supported by the U. S. Department of Energy, Office of Science, Office of Basic Energy Sciences, under Contract No. DE-AC02-06CH11357.

## APPENDIX

For the x-ray diffraction measurements, we have also performed spatial scans of the sample with the detector angle set at $29.92^\circ$. This enabled us to confirm that the monoclinic $M_1$ phase indeed transforms to the rutile phase. Fig. 5 displays complementary spatial images obtained at $T = 339.2$ K with the detector angles set at $29.58^\circ$ and $29.92^\circ$. The images are anti-correlated as expected.

In Fig. 6, we plot the first derivatives with respect to temperature of the monoclinic $M_1$ and insulating fractions shown in Fig. 4. Notice that the first derivative of the $M_1$ fraction becomes positive in the temperature regime where the $M_1$ fraction is non-monotonic. The insulating fraction never shows a positive first derivative because its temperature dependence is always monotonic.

Over 150 scans of the Bragg intensity as a function of detector angle $2\theta$ were performed at various temperatures in the phase transition regime of $VO_2$. The scans were performed on various 160 nm x 40 nm size regions (our spatial resolution) located within the 2 μm x 2 μm zone of interest. All such scans can be fitted to a linear combination of the $M_1$ and rutile Bragg peaks. A typical example is shown in Fig. 7. Bragg intensity within the scanned detector angles from a hypothetical intermediate phase cannot be larger than the range of the residuals shown in Fig. 7. This fact argues against the presence of any significant fraction of an intermediate lattice structure within the phase transition regime in the (200) oriented $VO_2$ films grown on $(\bar{1}012)$ sapphire.

Both x-ray nano-diffraction and near-field infrared microscopy are bulk probes that are not sensitive to surface effects. One difference between them, however, is the depth probed by the two techniques. While x-rays penetrate the entire depth of the 80 nm film, SNIM is mostly sensitive up to ~ 50 nm depth from the film surface[35]. Therefore, it may be that the non-monotonic switching behaviour of the structural transition occurs in the vicinity of the film-substrate interface, a region that is difficult to access by SNIM. However, there is no obvious reason why this phenomenon would be limited to the interface of the film with the substrate. Moreover, the vacillation of the structure appears unlikely to be merely an interface phenomenon because it occurs in a regime when a significant fraction of the bulk of the film has gone through the SPT. If we consider the possibility that the interface of the film is at a slightly higher temperature than the surface, then on thermodynamic grounds it is much less feasible for the high-temperature structure near the interface to return to the lower-temperature structure. Furthermore, the monotonic decrease of the resistance, which is sensitive to the entire thickness of the $VO_2$ channel, is in agreement with the monotonic decrease of the insulating fraction obtained from SNIM images. We also note here that the monotonic change in our resistance measurements is consistent with other reports on charge transport in micrometer-size $VO_2$ two-terminal structures. Such structures, which ought to be sensitive to small vacillations of the resistance, do not show any non-monotonic behaviour in the electronic transport. Hence, we do not think that the difference in the depth of the film probed by the two imaging techniques is the true reason behind the observed phenomena. Rather, our observations can be explained by nanoscale dichotomy between the electronic and structural changes.

* Permanent address: Department of Physics, College of William and Mary, Williamsburg, Virginia 23187, USA. Electronic mail: mumtaz@wm.edu

**Figure legends**

Fig. 1: Bragg intensity from the $M_1$ monoclinic phase of $VO_2$ obtained at $T$ = 330 K and the rutile phase of $VO_2$ obtained at $T$ = 360 K. The dashed vertical lines depict the detector angles of 29.58° and 29.92° that were fixed for scanning x-ray nano-imaging. The inset shows an optical image of the $VO_2$-based device that we studied with nano-imaging and transport measurements. Both the length and the width of the $VO_2$ channel between the tapered ends of the gold electrodes are 3.5 μm.

Fig. 2: Spatial maps of the same 1.2 μm x 1.7 μm area showing the variation in the intensity of the Bragg peak from the $M_1$ phase of $VO_2$ obtained with the detector angle set at 29.58°. These images were obtained while heating $VO_2$ through the SPT. The color scale in relative units is the same for all the images. Higher intensities (dark blue) indicate the presence of the $M_1$ phase, lower intensities (red) represent the rutile phase, and intermediate intensities (green color) indicates coexistence of $M_1$ and rutile phases within the 160 nm x 40 nm x 80 nm volume of the x-ray beam footprint.

Fig. 3: Spatial maps of the same 1.7 μm x 1.9 μm area showing the variation in the third harmonic of the infrared scattering amplitude. These images were obtained while heating $VO_2$ through the IMT. The color scale in relative units is the same for all the images. Higher amplitudes (light blue and white) are from the metallic regions while the lower amplitudes (dark blue) are from insulating regions.

Fig. 4: The fraction of monoclinic $M_1$ phase and the fraction of the insulating phase are plotted as a function of temperature along with the resistance of $VO_2$ in the SPT/IMT regime.

Fig. 5: Spatial maps of the same 2 μm x 2 μm area at $T$ = 339.8 K showing the variation in the intensity of the Bragg peak from (a) the $M_1$ phase of $VO_2$ obtained with the detector angle set at $2\theta$ = 29.58° (b) the rutile phase of $VO_2$ obtained with the detector angle set at 29.92°. The color scale in relative units is the same for both images and is identical to the color scale used for Fig. 2.

Fig. 6: Plots of the first derivative with respect to temperature of the monoclinic $M_1$ and insulating fractions shown in Fig. 4 of the manuscript.

Fig. 7: Bragg intensity is plotted as a function of the detector angle $2\theta$ at $T = 337.4$ K in the phase transition regime of $VO_2$. The data is fit to a linear combination of the monoclinic $M_1$ and rutile Bragg peaks. The difference between the data and the fit (the residual) is also plotted. The gray, hatched region represents the range for the residuals from numerous intensity versus $2\theta$ scans in the phase transition regime.

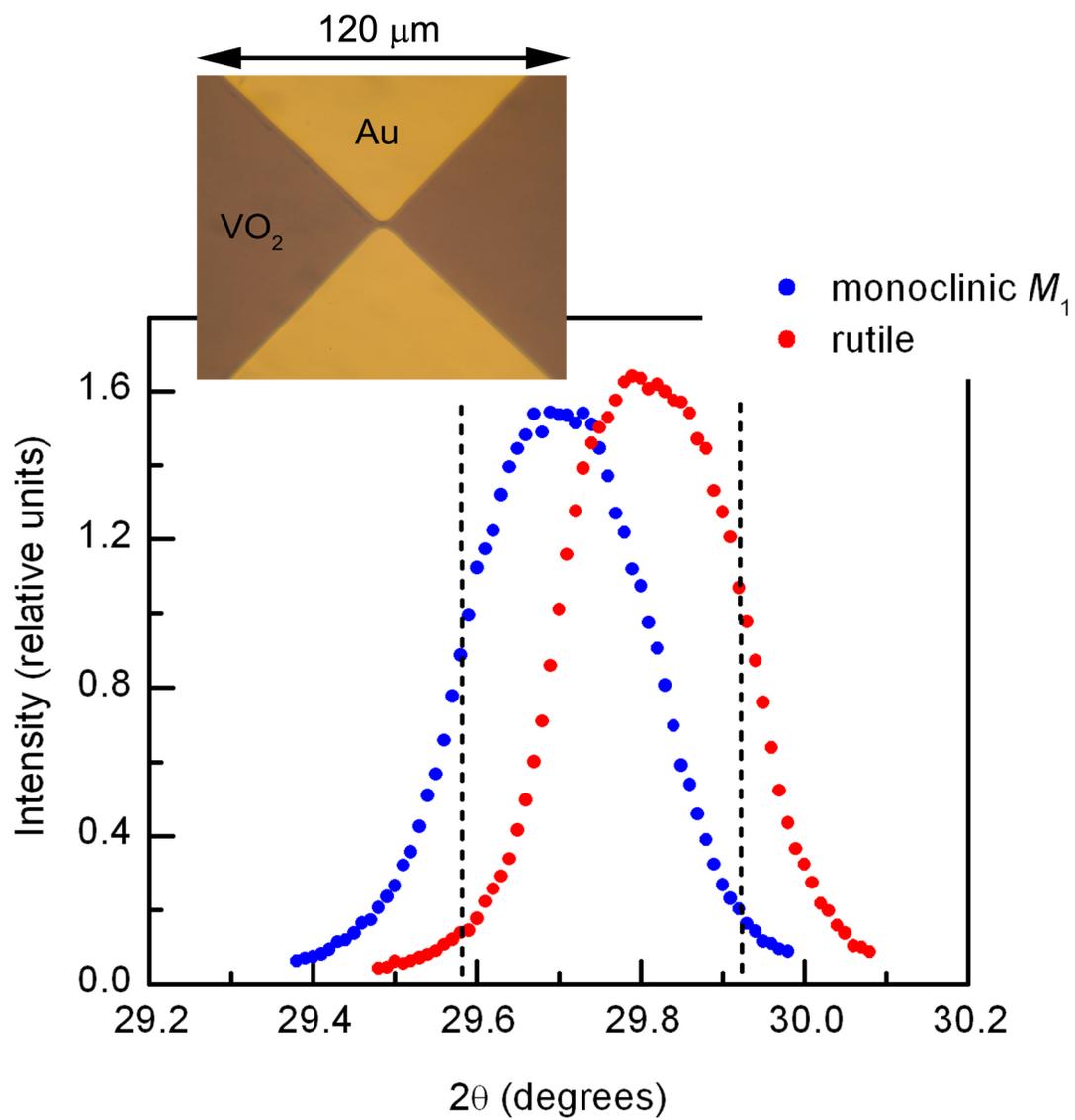

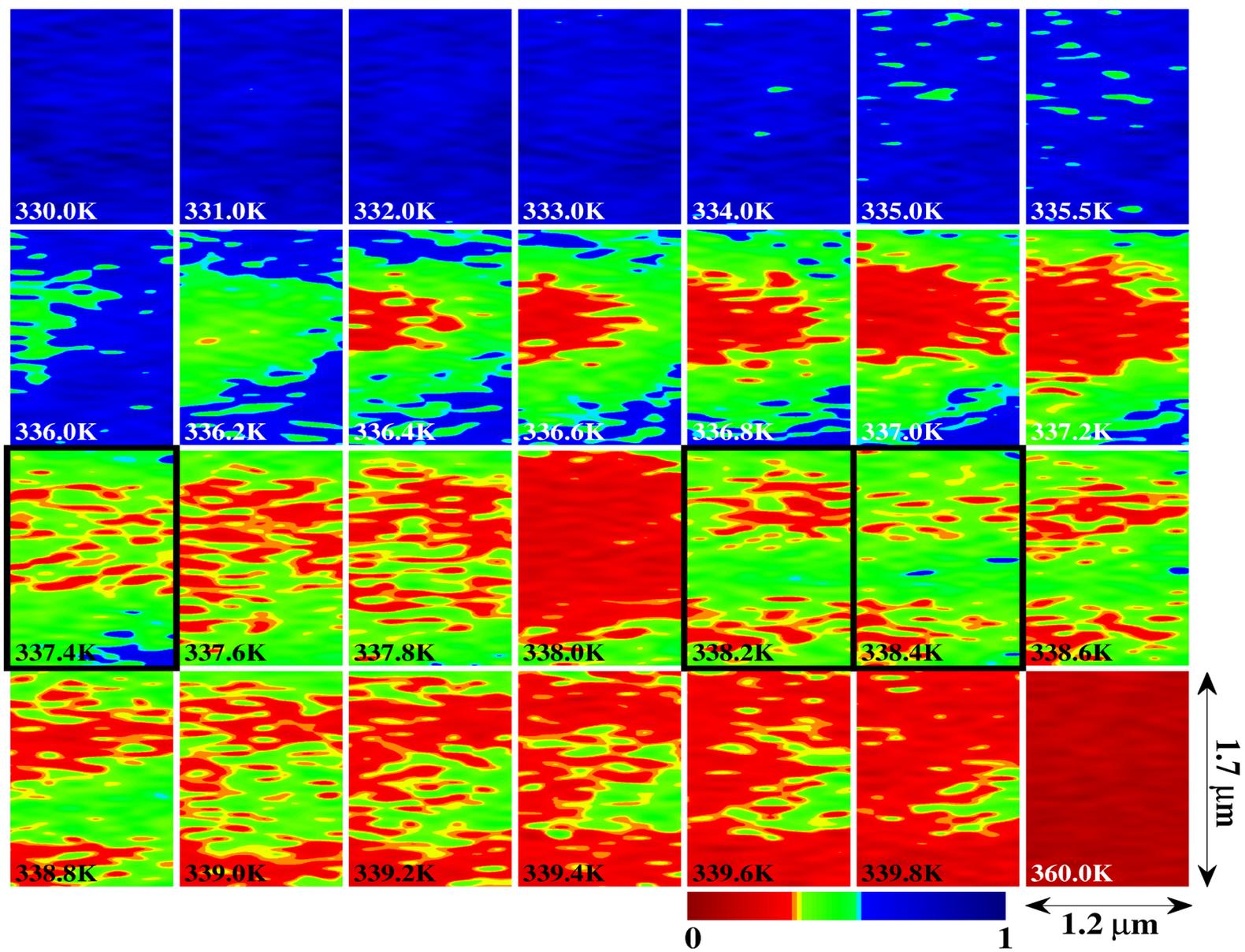

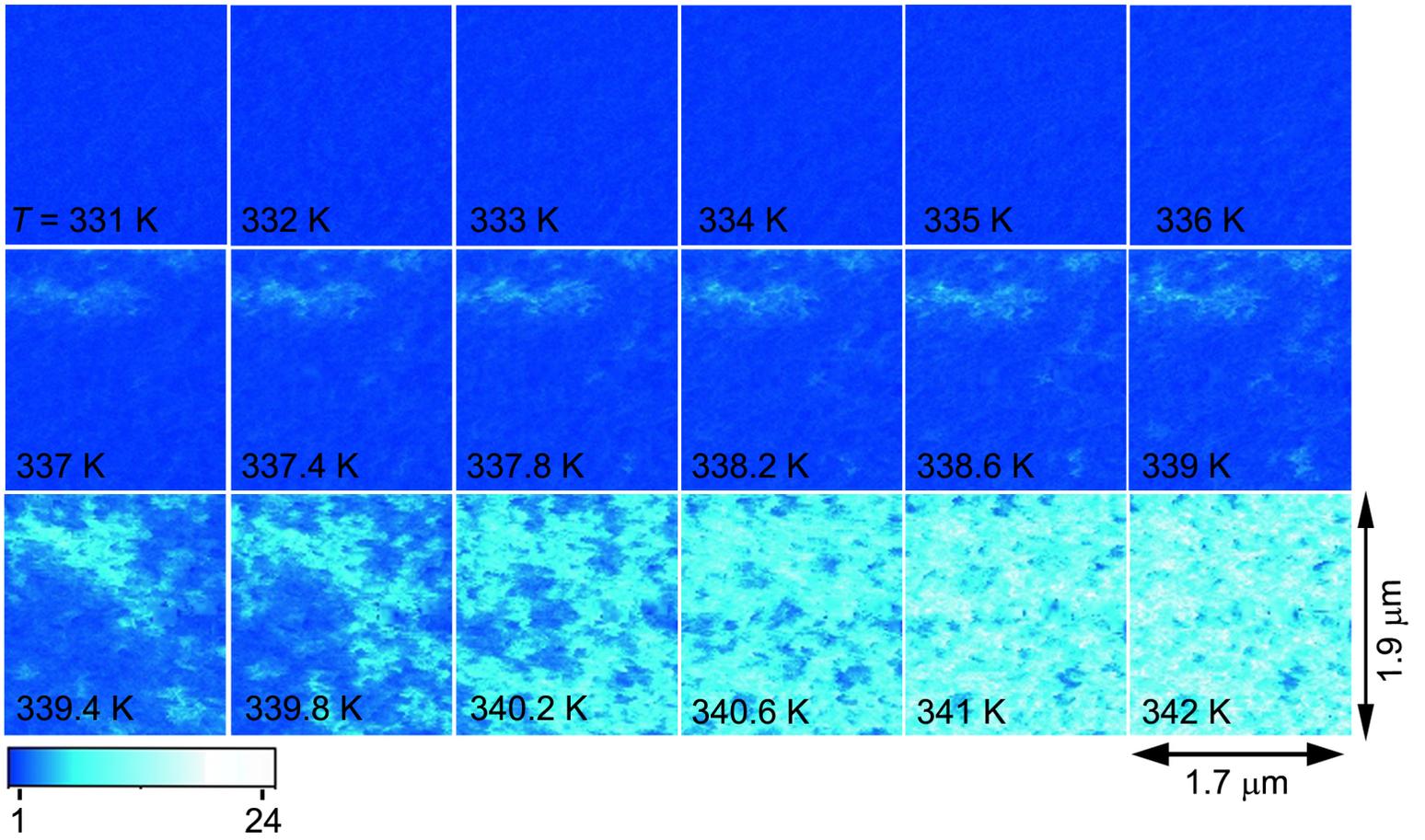

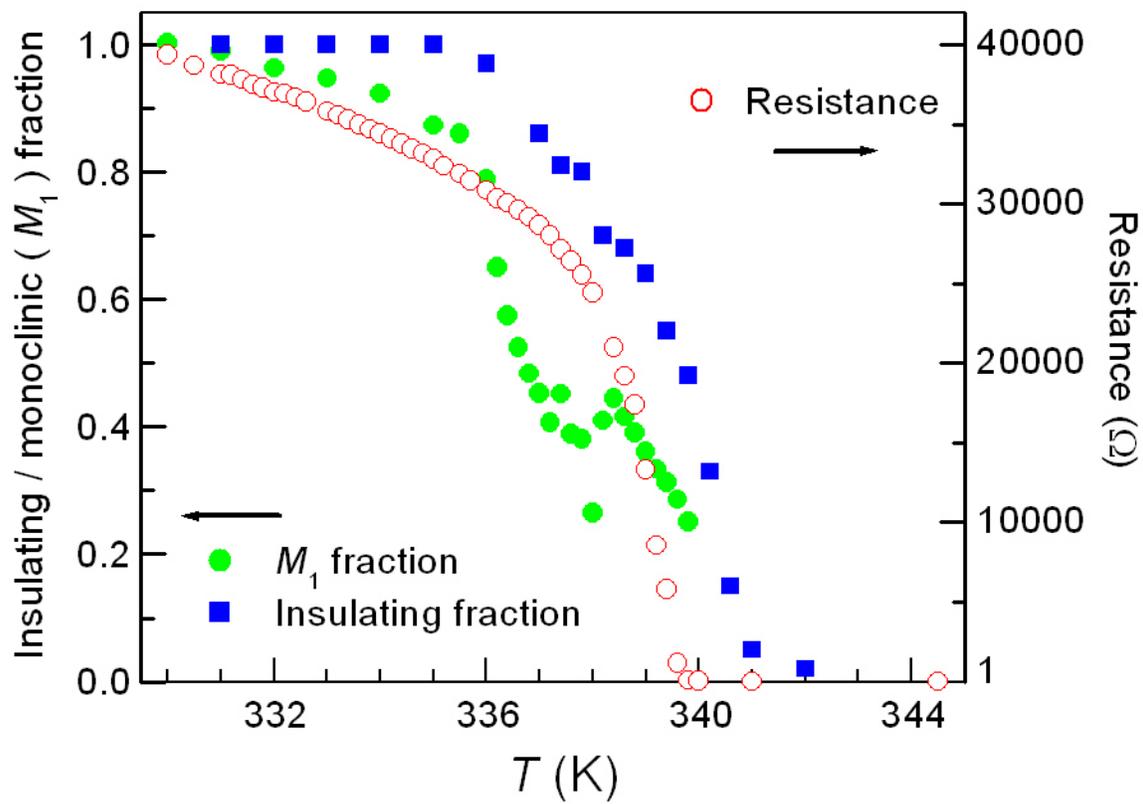

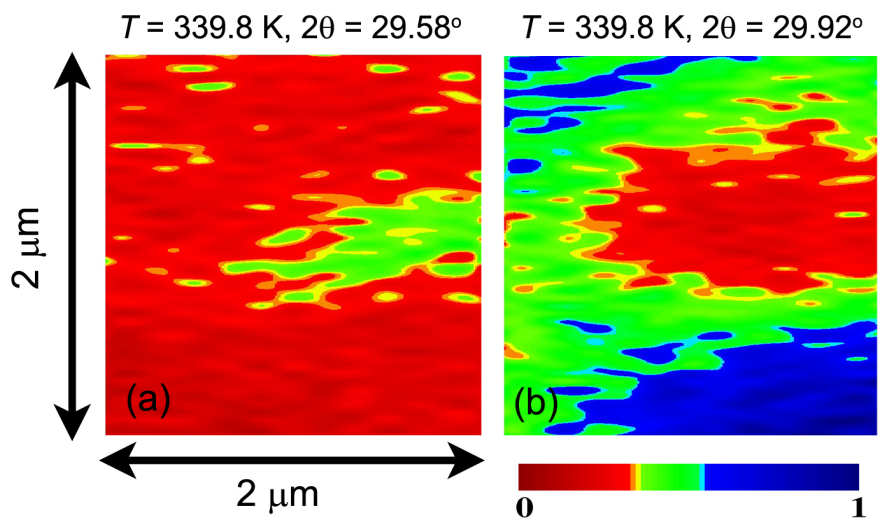

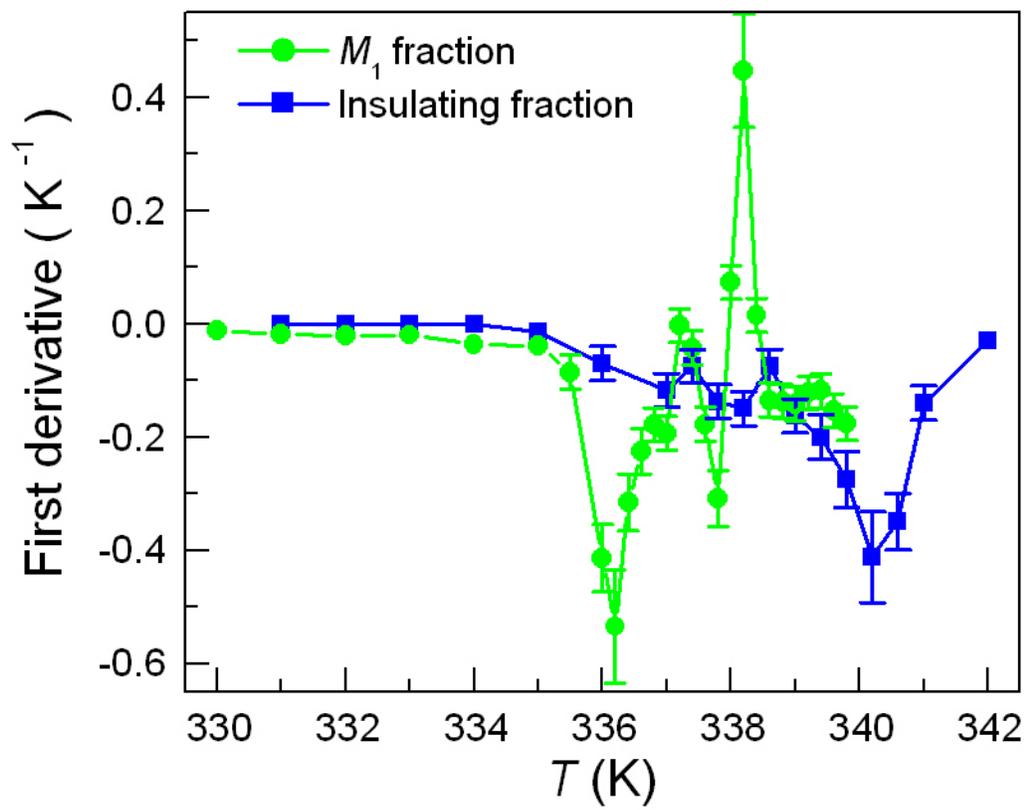

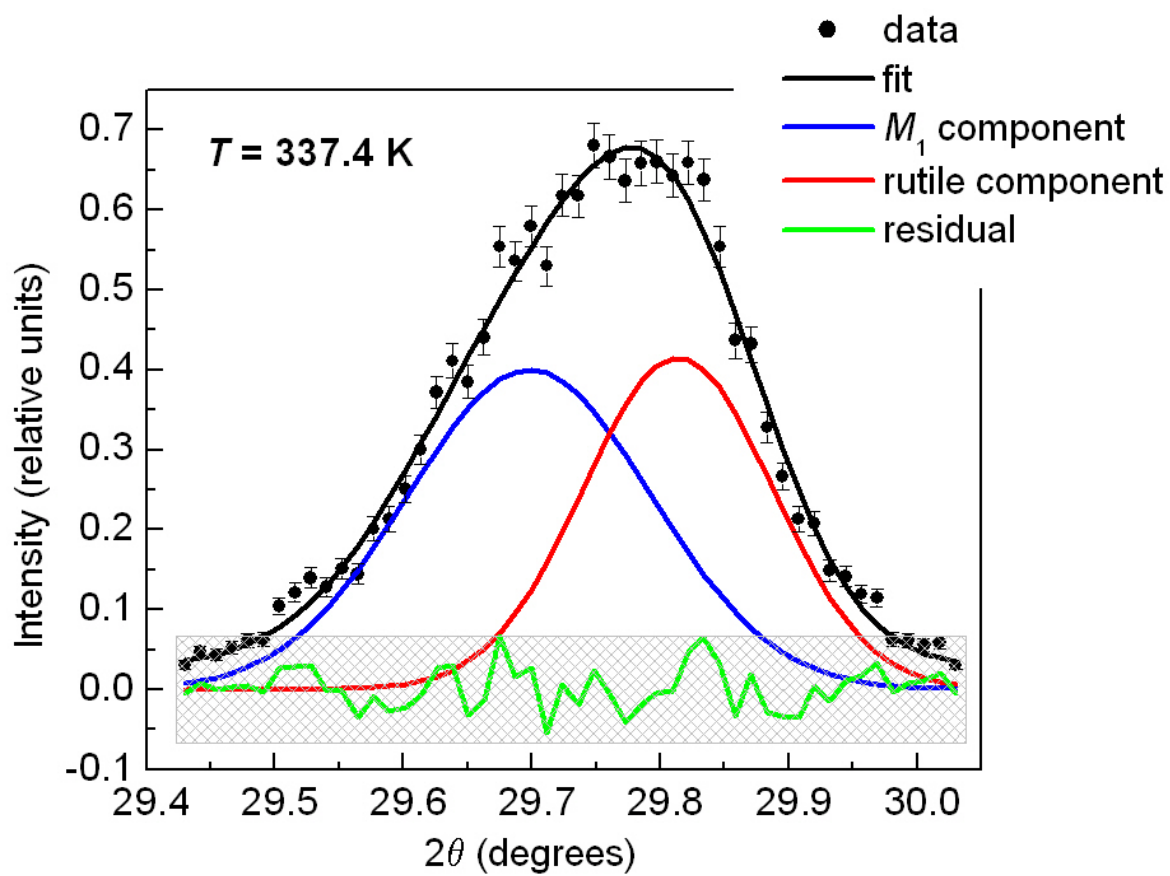